\documentclass[aps,prd,twocolumn]{revtex4}
\usepackage{exscale,latexsym,amsmath,amssymb} 
\usepackage[dvips]{graphicx}
\newcommand{\bk}[3]{\ensuremath{\left\langle #1 \right| #2  \left|  #3 \right\rangle}}
\newlength{\bredde}
\def\slash#1{\settowidth{\bredde}{$#1$}\ifmmode\,\raisebox{.15ex}{/}
\hspace*{-\bredde} #1\else$\,\raisebox{.15ex}{/}\hspace*{-\bredde} #1$\fi}
\begin{document}
\title{Leading quantum gravitational corrections to QED}
\author{\sc{M.~S.~Butt}}
\email[]{butmu@nbi.dk} 
\affiliation{The Niels Bohr Institute\\
Blegdamsvej 17, DK-2100 Copenhagen\\ Denmark}
\begin{abstract}\label{sec:abstract}
We consider the leading post-Newtonian and quantum corrections to the
non-relativistic scattering amplitude of charged
spin-{\mbox{\small$\frac{1}{2}$}} fermions in the combined theory of
general relativity and QED. The coupled Dirac-Einstein system is
treated as an effective field theory. This allows for a consistent
quantization of the gravitational field. The appropriate vertex rules
are extracted from the action, and the non-analytic contributions to
the 1-loop scattering matrix are calculated in the non-relativistic
limit. The non-analytical parts of the scattering amplitude are known
to give the long range, low energy, leading quantum corrections, are
used to construct the leading post-Newtonian and quantum corrections
to the two-particle non-relativistic scattering matrix potential for
two massive fermions with electric charge.
\end{abstract}
\maketitle
\section{Introduction}\label{sec:introduction} 
On the search for a theory of quantum gravity,
Donoghue~\cite{Donoghue1} proposed 12 years ago an interesting new way
to look at general relativity. He suggested that when treating general
relativity as an effective field theory~\cite{Weinberg3}, reliable
quantum predictions at the low energies could be made, in the same way
as chiral perturbation theory is used as the low energy approximation
of QCD, being the effective field theory of QCD. It is well known that
a field theory need not be strictly renormalizable in order to be able
to yield quantum predictions at low energies. A fundamental quantum
theory of gravity does not appear in this way, but it is possible to
calculate quantum corrections order by order in a momentum expansion.

Having laid the foundations for this new approach, Donoghue and
collaborators turned their attention to the practical applications of
this idea. A number of interesting calculations has been made
involving quantum gravitational corrections to various
quantities~\cite{Donoghue,Donoghue1,Bohr1,Bohr2,Holstein}.

Prior to the effective field theoretical description of general
relativity, attempts had been made to find a quantum theory for
gravity. In particular many proved that general relativity was indeed
not renormalizable, be it pure general relativity or general
relativity coupled to bosonic or fermionic matter, see
e.g.~\cite{Veltman,'tHooft,Deser,Gast}. Of course it is a well known
fact that general relativity is a \emph{non}-renormalizable theory per
se, and these authors succeeded in exactly confirming that gravity
indeed is explicitly non-renormalizable, with or without
matter. However, when looked at in the framework of an effective field
theory, these theories do become order by order renormalizable in the
low energy limit. All possible counter-terms, also those not present
in the initial Lagrangian, are generated. When general relativity is
treated as an effective theory, renormalizability simply fails to be
an issue. The ultraviolet divergences arising e.g. at the 1-loop level
are dealt with by renormalizing the parameters of higher derivative
terms in the action. Many interesting results have been found from
this procedure. Most interesting for the point of view of this paper
is the bosonic quantum corrections to the Newtonian/Coulomb
potentials~\cite{Bohr1}.

When approaching general relativity in this manner, it is convenient
to use the background field method~\cite{Dewitt,Abbott}. Divergent
terms are absorbed away into phenomenological constants which
characterize the effective action of the theory. The price paid is the
introduction of a set of never-ending higher order derivative
couplings into the theory. General relativity thus turns into a
minimal theory in the myriad of higher order terms in the action,
however still remaining a valid theory for gravitational interactions
at low energies. The effective action contains all terms consistent
with the underlying symmetries of the theory. Perturbatively certain
terms of the action play leading roles at certain energy scales, hence
only a finite number of terms are required to be accounted for at each
loop order. Here we consider the low energy limit of the effective
theory of quantum gravity. Because we work to lowest order, our
results are free of the new additional terms that must be appended to
the Einstein action and which manifest themselves as components of the
high energy couplings of the effective field theory.

In ref.~\cite{Bohr1} the post-Newtonian as well as the quantum
corrections, that were generated to the Newtonian and Coulomb
potentials were explicitly found. We wish to repeat this calculation,
but now in terms of couplings to fermions. We wish in particular to
see explicitly if the post-Newtonian as well as the quantum
gravitational corrections generated are identical to~\cite{Bohr1} or
not.

We will more or less follow a similar procedure as in~\cite{Bohr1}
mostly to avoid confusion about conventions and to make it easy to
compare the results at the end. However, it is by no means a
straightforward task to complete a similar calculation in terms of
fermions. Some obstacles have to be overcome compared to bosonic
matter in curved space, e.g. the issue of introducing fermionic matter
into curved space-time. Luckily, this issue has been dealt with
before~\cite{Utiyama,Weyl,Cartan,Birrell,Deser,Gast,Milton}. The
additional formalism required is the introduction of the vierbein
formalism and deriving a proper covariant derivative for the spinor
fields.

Donoghue devised a particular elegant way to extract relevant
information in terms of analytical and non-analytical contributions to
the scattering matrix. This was realized through the integrals
occurring in the calculations, and propagation of massless
particles. Because the post-Newtonian and the quantum corrections are
fully determined by the non-analytical pieces of the 1-loop amplitude
generated by the lowest order Einstein action, it becomes possible to
perform this calculation completely. We will also only consider 1-loop
effects in this paper. We will extract the non-analytical parts of the
full set of 1-loop diagrams needed for the 1-loop scattering matrix in
the combined quantum theory of general relativity and QED. As we will
see in this paper, and as can also seen
in~\cite{Donoghue,Donoghue1,Bohr1,Bohr2,Holstein}, the non-analytical
contributions correspond exactly to the long range corrections of the
potential.

We will employ the same conventions as in
ref.~\cite{Bohr1,Donoghue}. The mostly minus Minkowski metric
convention $(1,-1,-1,-1)$ will be used and the natural units are
$(\hbar=c=1)$ when nothing else is stated.

In section~\ref{sec:intr-vierb-fields} we will shortly review the
concept of vierbein fields and how to introduce a proper covariant
derivative for the spinors when working with fermions. We will also
quantize the metric and vierbeins using the background field
method. We will explain the correspondence between the metric and
vierbein formulation of our theory. In
section~\ref{sec:dirac-einst-syst} we will see how to combine QED with
general relativity by using the vierbein formalism and moreover
introduce the ghost fields. Next in section~\ref{sec:scatt-matr-thep}
we will focus first on the distinction between non-analytical and
analytical contributions to the scattering matrix amplitude,
where after we will define the potential. Finally, in the succeeding
section~\ref{sec:results-feym-diagr}, we will evaluate the Feynman
diagrams contributing non-analytically to the scattering matrix, in
order to construct the leading corrections to the non-relativistic
Newtonian and Coulomb potential. We will end this paper with a
discussion in relation to~\cite{Bohr1}. In the appendix, the vertex
rules are presented.
\section{Introducing the vierbein fields}\label{sec:intr-vierb-fields}
At every space-time point $x_{0}$ it is possible to erect a set of
coordinates $\xi^{a}_{x_{0}}$ locally inertial at the given point in
question, in accordance with the theory of special relativity. This in
turn implies that the erected set of coordinates $\xi^{a}_{x_{0}}$
vary from point to point, and that the information about the
gravitational field is in fact contained in the change of the local
inertial coordinate systems from point to point. Hence it is possible
to express $\xi^{a}_{x_{0}}$ as a local function of any non-inertial
coordinates (i.e. in a general coordinate system) $x^{\mu}$ i.e.
\begin{subequations}
\begin{align}\label{eq:1}
d\xi^{a} &=e^{a}{}_{\mu} dx^\mu \\
dx^{\mu} &=e_{a}{}^{\mu} d\xi^a
\end{align}
\end{subequations}
evaluating the derivatives at the point of interest, where the inverse
operation is also given at the point in question. The transformation
matrix relating the local inertial frames to the arbitrary coordinate
system is the called vierbein field, it is denoted $e^{a}{}_{\mu}(x)$
and is a function of $x^{\mu}$.

It is seen that due to the transformations $x \rightarrow \xi$ and
$\xi \rightarrow x$ being nonsingular, $e_{a}{}^{\mu}(x)$ is the
inverse transformation matrix.

One can find other relations between the vierbein fields, e.g
\begin{subequations}
\begin{align}\label{eq:2}
 e^{a}{}_{\mu} e_{b}{}^{\mu} &= \delta^{a}_{b}\\
 e_{a}{}^{\mu} e^{a}{}_{\nu} &= \delta^{\mu}_{\nu}
\end{align}
\end{subequations}
The metric expressed in general coordinates can be found by looking at the
proper time in general coordinates
\begin{align}\label{eq:3}
 d\tau^{2} &= \eta_{ab} d\xi^{a}(x)d\xi^{b}(x)
\end{align}
where the metric in general coordinates is
\begin{align}\label{eq:4}
 g_{\mu \nu}(x) &=e^{a}{}_{\mu}(x)e^{b}{}_{\nu}(x) \eta_{ab}
\end{align}
where $a,b\ldots$ are Lorentz indices and $\mu ,\nu \ldots$ are the
general coordinate indices. The inverse of the metric is
\begin{align}\label{eq:5}
 g^{\mu \nu} &=e_{a}{}^{\mu}(x)e_{b}{}^{\nu}(x) \eta^{ab}
\end{align}
from which a familiar result is obtained
\begin{align}\label{eq:6}
\nonumber g_{\mu \nu} g^{\nu \sigma} &= \big( e^{a}{}_{\mu}e^{b}{}_{\nu}
\eta_{ab} \big) \big( e_{c}{}^{\nu}e_{d}{}^{\sigma} \eta^{cd} \big)
= e^{a}{}_{\mu} e_{a}{}^{\sigma} = \delta^{\sigma}_{\mu}
\end{align}
Thus the $e^{a}{}_{\mu}(x)$ fields relate the Lorentz axes to the
coordinate axes at each point in space-time, explicitly
\begin{equation}\label{eq:7}
e^{a}{}_{\mu}(x=x_{0})= \left( \frac{ \partial \xi^{a}_{x_{0}}(x)
}{\partial x^{\mu}} \right)_{x=x_{0}}
\end{equation}
When changing the general coordinates from $x^{\mu}\rightarrow
{x'}^{\mu}$, these fields transform as covariant vectors and under
local Lorentz transformations they transform as Lorentz vectors
\begin{subequations} 
\begin{align}
\label{eq:8} e^{a}{}_{\mu}(x) &\rightarrow e^{\prime a}{}_{\mu}(x') = \frac{\partial
x^{\nu}}{\partial {x'}^{\mu }} e^{a}{}_{\nu}(x)\\ 
\label{eq:71} e^{a}{}_{\mu} &\rightarrow e'^{a}{}_{\mu} =  \Lambda^{a}{}_{b}
e^{b}{}_{\mu}
\end{align}
\end{subequations}
The vierbein fields are the only index changing objects in this
theory. For given covariant/contravariant vector fields or also a
tensor field, we can refer their components at $x$ to the locally
inertial coordinate system $\xi^{a}_{x_{0}}(x)$ at $x_{0}$ by using
the vierbein
\begin{align}\label{eq:9}
 e^{a}{}_{\mu} A^{\mu}&= A^{a}
\end{align}
$A^{a}$ transforms as a collection of four scalars under general
coordinate transformations \eqref{eq:8}, and under the local Lorentz
transformations \eqref{eq:71} it behaves as a vector. It is thus
possible to use the vierbeins, to convert general tensors into local,
Lorentz-transforming tensors, whereby shifting the additional
space-time dependence into the vierbeins.
\subsection{Gauge transformation of the fermion fields and the spin
connection }\label{sec:gauge-transf-ferm}
Using the generators of the Lorentz group
$\sigma_{ab}=\frac{1}{4}[\gamma_{a},\gamma_{b}]$ for spinors belonging
to a given representation $S(\Lambda)=
e^{\frac{1}{2}\lambda^{ab}\sigma_{ab}}$ for small $\lambda$ in infinitesimal form we get
for spinor transformations
\begin{equation}\label{eq:10} 
\psi ' \rightarrow \big( 1 + \frac{1}{2} \sigma_{ab} \lambda^{ab}(x)
\big) \psi
\end{equation}
Since partial derivatives on fields almost always occur in the
Lagrangian density functions, it would be worthwhile to subject
\eqref{eq:10} to a partial differentiation, and see if it
transforms covariantly or not. We immediately see from the
transformation
\begin{equation}\label{eq:11}
\partial_{\mu} \psi ' \rightarrow \big(1 + \frac{1}{2} \sigma_{ab}
\lambda^{ab}(x) \big) \partial_{\mu}\psi + \Big[\frac{1}{2}
\sigma_{ab} \partial_{\mu } \lambda^{ab}(x)\Big] \psi
\end{equation}
that the fermion field does not transform as a proper Lorentz spinor
under this operation, even though the ordinary derivative
$\partial_{\mu} \psi$ is a covariant vector since the fermion fields
are defined to be scalar objects under coordinate transformations. In
order to get a feasible theory \eqref{eq:11} must transform
covariantly, hence we must invent a "covariant derivative" $D_{\mu}$
for fields that transform in this manner, such that in the end result
the second term in \eqref{eq:11} gets canceled.

We can find this object by studying the behavior of the vierbein
fields under the transformations \eqref{eq:8} and \eqref{eq:71} and
comparing it with the transformation properties of a general four
vector~\cite{Veltman}.

The rule for covariant derivative for an object mixed in both Lorentz
and general coordinate indices $A^{a}_{\mu}$ is~\cite{Thesis}
\begin{equation}\label{eq:12}
D_{\mu}A^{a}_{\nu} = \partial_{\mu} A^{a}_{\nu} -
\Gamma^{\alpha}_{\mu \nu} A^{a}_{\alpha} + \omega_{\mu }{}^{a}{}_{
b}A^{b}_{\nu}
\end{equation}
where $\omega_{\mu }{}^{a}{}_{ b}$ is called the spin connection. To
determine the structure of the spin connection one can require that
the covariant differentiation should commute with the operation of
index changing, not only just index lowering/raising~\cite{Deser} ,
thus using the condition \footnote{This is in analogy with the affine connection relation
$\Gamma_{\mu \nu}^{\lambda}$ that followed from $D_{\alpha}g_{\mu
\nu}=g_{\mu \nu ,\alpha}= 0 $ }
\begin{equation}\label{eq:13}
D_{\nu}e^a_{\ \mu} = \partial_\nu e^a_{\ \mu} - \Gamma^\sigma_{\mu\nu}
e^a_{\ \sigma} + {\omega_{\nu}}_{\ b}^a e^b_{\ \mu} = 0
\end{equation}
one finds that the following object does the job
\begin{equation}\label{eq:14}
\begin{aligned} 
\omega_{\mu}^{ab} &= \frac{1}{2} \Big(
e^{[a\nu}\partial_{[\mu}e^{b]}{}_{\nu]} + e^{a\rho} e^{b\sigma}
\partial_{[\sigma} e_{c\rho]}e^{c}{}_{\mu } \Big)
\end{aligned}
\end{equation}
Note that the indices commute only with alike (space-time-)indices,
i.e. Latin characters with Latins and Greek characters with Greek.

This object has exactly the transformation property that we were
searching for
\begin{equation}\label{eq:15}
\omega_{\mu}^{ab} \rightarrow \omega'^{ab}_{\mu} - \partial_{\mu}
\lambda^{ab}
\end{equation}
where the primed object shows transformation as a tensor. The
covariant derivative for spinor fields now becomes
\begin{equation}\label{eq:16}
D_{\mu} \psi \equiv (\partial_{\mu} + \frac{1}{2} \sigma_{ab} \omega_{\mu}^{ab})\psi 
\end{equation}
This will transform as a proper Lorentz spinor and a covariant vector under
the mentioned gauge transformations. 
\subsection{Quantization of the vierbein and metric fields}\label{sec:quant-vierb-metr}
To quantize the combined theory of QED and gravity we will have to
quantize the vierbein fields in the same way as it is done when
quantizing the metric.

In the background field method the quantum corrections to general
relativity are described by quantum vibrations of the metric tensor,
making it possible to expand the metric and vierbein fields into two
separate contributions, a classical background field and a quantum
field
\begin{align}\label{eq:17}
g_{ \mu \nu } &= \bar g_{ \mu \nu } + \kappa h_{ \mu \nu }\\
  e^{a}_{~\mu} &= \bar e^{a}_{~\mu} +\kappa c^{a}_{~\mu} 
\end{align}
where $\kappa^{2}=32 \pi G$ and the background fields are denoted as
$\bar g_{ \mu \nu }$ and $\bar e^{a}_{~\mu}$. The quantum part - the
graviton field - is denoted by $ h_{ \mu \nu }$ and $ c^{a}_{~\mu} $, the
sum of these being the full metric and vierbein respectively. We will
soon see how the metric and vierbein quantum variables for the
gravitons are related. Furthermore we find the following inverses and
other relations
\begin{equation}\label{eq:18}
\begin{aligned}
e_{a \mu}  &= \bar e_{a \mu} +  \kappa c_{a \mu}\\
e^{a\mu}     &= \bar {e}^{a\mu}      - \kappa c^{\mu a} + \ldots \\
e_{a}^{~\mu} &= \bar {e}_{a}^{~\mu} - \kappa c^{\mu}_{~a} 
+ \ldots  \\
\end{aligned}
\end{equation}
for the vierbeins
\begin{equation}\label{eq:19}
\begin{aligned}
g_{ \mu \nu }    &= \bar g_{ \mu \nu } +  \kappa h_{ \mu \nu } \\
&= \bar g_{\mu\nu} + \kappa ( c_{\mu\nu} +c_{\nu\mu}  ) + \kappa^{2} c_{a\mu}c^{a}{}_{\nu}\\
g^{ \mu\nu } &= \bar g^{ \mu \nu } - \kappa h^{\nu \mu }  +\ldots\\
 &= \bar g^{\mu \nu} - \kappa ( c^{\mu\nu} +c^{\nu\mu} ) + \ldots
\end{aligned}
\end{equation}
for the metric. We have only expanded the inverses to first order in
the quantum fields, as we need diagrams to second order in
$\kappa$. From these relations we see that the metric and vierbein
quantum fields are related according to
\begin{equation}\label{eq:20}
h_{\mu\nu} = c_{\mu\nu} +c_{\nu\mu} + \kappa c_{a\mu}c^{a}{}_{\nu} =
s_{\mu \nu } + \kappa c_{a\mu}c^{a}{}_{\nu}
\end{equation}
showing us that the quantized metric field is equal the quantized
symmetric vierbein field to first order in the quantum fields,
i.e. $h_{\mu\nu} = c_{\mu\nu} +c_{\nu\mu}\equiv s_{\mu\nu}$. It is
easy to deduce the determinants of the vierbein and metric fields
\begin{equation}\label{eq:21}
 e= \det[e^{a}_{~\mu}] \approx \tilde{e} \Big( 1+ \kappa
c^{\alpha}_{~\alpha} + \ldots\Big)
\end{equation}
with $ \tilde{e} \equiv \det \bar {e}^{a}_{~\alpha}$, as well as the
square-root of the metric tensor $ \sqrt{ - g }= \sqrt{- {\rm det}(g_{\mu\nu})}$
\begin{equation}\label{eq:22}
\sqrt{-g} \approx \sqrt { - \bar
g } \Big( 1 + \frac{\kappa}{2} h^{ \alpha }_{ \alpha } + \ldots
\Big)
\end{equation}
with $ \bar g \equiv \det \bar {g}_{\mu\alpha}$. Finally it will
be necessary to expand the spin connection in terms of the vierbein
fields as well, the leading order terms appearing in the calculations
are (in flat background field)
\begin{subequations}
\begin{align}\label{eq:23}
  w_{\mu ab}^{{\rm Background}}        &=   0  \\
  w_{\mu ab}^{{\rm First\ order} } &= \frac{1}{2} \partial_\mu a_{ba}
+ \frac{1}{2}\partial_b s_{a\mu} - \frac{1}{2}\partial_a s_{b \mu}
\end{align}
\end{subequations}
where we have defined a new field, an antisymmetric field $a_{\mu\nu}=c_{\mu\nu}-c_{\nu\mu}$.
\section{The Dirac-Einstein system as a combined effective field theory}\label{sec:dirac-einst-syst}
The combined theory of QED in a gravitational field is given by the
sum of the QED and Einstein Lagrangian densities
\begin{equation}\label{eq:24}
\mathcal{L} = \mathcal{L}_{\rm Gravity} + \mathcal{L}_{\rm QED}
\end{equation}
The interacting field theory for Quantum Electrodynamics is well
known, with the Dirac equation minimally coupled to the
electromagnetic field
\begin{equation}\label{eq:25}
\begin{aligned}
\mathcal{L}_{\rm QED}
&= \mathcal{L}_{\rm Dirac} + \mathcal{L}_{\rm Maxwell}\\
&= \bar \psi (i \gamma^{ \mu } D_{\mu}   -m ) \psi
-\frac{1}{4} g^{ \alpha \mu } g^{ \beta \nu } F_{ \alpha \nu } F_{ \mu
\beta }
\end{aligned}
\end{equation}
where $m$ is the mass, $e_{q}$~\footnote{The $q$ is attached to
distinguish the charges from the vierbeins.} is the electron charge
with $e_{q}=|e_{q}|$ and finally $D_{\mu}\equiv \partial_{\mu} - i
e_{q} A_{\mu}(x)$ is the covariant derivative.

To make the action of the Dirac Lagrangian density invariant under
general coordinate transformations, we follow the general procedure,
i.e. multiply it with $\sqrt{-g}$ and at the same time introduce our
new covariant derivative
\begin{equation}\label{eq:26}
 \mathcal{L}_{\rm Dirac} = \sqrt{ - g } \bar \psi \left(i
\gamma^{ \mu } D_{ \mu }-m \right) \psi 
              =  e \bar \psi\left(i \gamma^{ d } e_{ d }^{~ \mu }
D_{\mu} -m \right) \psi
\end{equation}
now $D_{\mu}= \partial_{ \mu } -i e_q A_{ \mu } + \frac{1}{2} \sigma^{
ab } w_{ \mu ab }$ and we have used $ \sqrt{ - g } =
{\rm det}(e^{a}{}_{\mu}) \equiv e$ i.e. the determinant of the vierbein is
the matrix square-root of the metric. Finally $ \gamma^{ \mu } =
\gamma^{ a } e_{ a }^{~ \mu } $.

The full generally covariant Lagrangian density including the
fermionic degrees of freedom may collectively be written as
\begin{multline}\label{eq:27}
\mathcal{L} = e \frac{2}{\kappa^{2}} R + e ( \bar \psi i
\gamma^{a}e_{a}{}^{\mu}D_{\mu}\psi - \bar \psi \psi m )\\
 -\frac{1}{4}\sqrt{-g} g^{ \alpha \mu } g^{ \beta \nu } F_{ \alpha \nu
} F_{ \mu \beta }
\end{multline}
This will account for our full theory. The Lagrangian density is to be
expanded in powers of $c_{\mu\nu}$ (where we choose $c_{\mu\nu}$ to be
linearly symmetrically equal to $h_{\mu\nu}$ as seen earlier) in the
case of the Dirac field and only $h_{\mu\nu}$ in the case of the
Maxwell fields, specifically we expand the Lagrangian density as follows
\begin{equation}\label{sec:dirac-einst-syst-2}
 \mathcal{L} = \mathcal{L}_{\rm Background} + \mathcal{L}_{\rm Linear\
order} +\ldots
\end{equation}
where the ellipses denote second and higher order terms that will not
contribute at the 1-loop level calculations.

The Lagrangian density for the photon field can now be expanded in
powers of $h_{\mu\nu}$, explicitly
\begin{widetext}
\begin{multline}\label{eq:28}
  \mathcal{L}_{\rm Maxwell} = - \frac\kappa4 h ( \partial_\mu A_\nu
\partial^\mu A^\nu - \partial_\nu A_\mu \partial^\mu A^\nu )
+ \frac\kappa2   h^{ \rho \sigma } ( \partial_{ \alpha } A_{ \sigma}\partial^{ \alpha } A_{ \rho}   +  \partial_{ \sigma} A_{ \alpha } \partial_{ \rho} A^{ \alpha }  
 - \partial_{ \alpha } A_{ \sigma}\partial_{ \rho} A^{ \alpha } -
\partial_{ \sigma} A_{ \alpha } \partial^{ \alpha } A_{ \rho} )
\end{multline}
where the trace of $h \equiv
h^\alpha_{~\alpha}=h_{\alpha}^{\alpha}=h$. And likewise for the
fermionic part
\begin{multline}\label{eq:29}
\mathcal{L}_{\rm Dirac}= e_{q}\bar \psi \gamma^{\mu} A_{\mu}\psi +
\frac{i\kappa e_q}{2} \bar{ \psi } \gamma^{ a } (
I_{a}^{~\mu\beta\alpha} -
\delta_{a}^{~\mu}\eta^{\alpha\beta} ) h_{\alpha\beta} A_{ \mu } \psi
 + \frac\kappa2 h ( \bar \psi i\gamma^{ \mu } \partial_{ \mu } \psi -
\bar \psi m \psi ) - \frac{\kappa}{2}\bar{ \psi } \gamma^{ d } h^{
\mu}_{ ~d } \partial_{ \mu } \psi
 + \frac{\kappa}{2}\bar{ \psi }
\gamma^{ \mu } \sigma^{ ab } \partial_b h_{a\mu} \psi
\end{multline}
\end{widetext}
where the symmetric identity $I_{a}{}^{\mu\beta\alpha} =\frac{1}{2}
\eta_a{}^{ \{\beta} \eta^{\alpha \}\mu}$.

All the necessary lowest order interaction vertices of fermions,
gravitons and photons can be found for the theory from the Linear
order expansions as stated above in equations \eqref{eq:28} and
\eqref{eq:29}. A summary of these rules is presented in the appendix.

In principle we should also expand
\begin{equation}\label{eq:30}
\mathcal{L} = e \frac{2}{\kappa^{2}} R
\end{equation}
however it is known~\cite{Deser} that the metric and vierbein
formulations are equivalent for fields with only covariant vector
indices. The coupling to the vierbein field only occur as symmetric
combinations of vierbein fields, the symmetric combination is as we
have seen to linear order equal to the metric tensor. No new aspects
of the traditional quantization of the pure gravitational action are
introduced. We will therefore use the known vertices and propagators
for the bosonic and gravitational fields.

We have excluded the antisymmetric fields in all our expressions. This
is due to the fact that the antisymmetric fields have propagators that
go as $\sim \kappa^2$. This can be seen when we fix the gauges in our
quantization scheme. As pointed out earlier, our theory (i.e.
fermions including gravitational effects) has two types of
invariances. One is the general coordinate
transformations~\eqref{eq:8}, under which the fermions behave as
scalars (since they are defined with respect to the local Lorentz
frame). The other is the local Lorentz transformations~\eqref{eq:71},
under which the fermions transform as spinors. If Einstein action is
included, then the coordinate gauge can be fixed by choosing the
harmonic (de Donder) gauge
\begin{equation}\label{eq:31}
\mathcal{L}^{C} = - \frac{1}{2} \sqrt{-\bar g} ( h_{\mu\nu}{}^{,\nu} -
\frac{1}{2} h_{\nu}{}^{\nu}{}_{,\mu} )^{2}
\end{equation} 
whereas the local Lorentz invariance is broken by
choosing the sum of the squares of the antisymmetric vierbein
components 
\begin{equation}\label{eq:32}
\mathcal{L}^{L} = -\frac{1}{2} e \kappa^{-2}a_{\mu
\nu}^{2}
\end{equation}
Gauge fixing of both these fields will result in an introduction of
two sets of ghost fields. We do not need to be concerned about the
ghost introduced due to the antisymmetric field. In a vierbein
description of pure gravity, the ghosts are never external,
furthermore neither the antisymmetric vierbein fields nor its ghosts
propagate (they cancel each other~\cite{Deser}), thus we will not need
to calculate vertices for the external ghosts fields. This is very
reassuring since the pure gravity theory in vierbein formulation can
be covariantly quantized and is equivalent to the quantized metric
approach. That is we could in principle describe the theory without
introducing these variables. But if we do not have pure gravity and
include fermions, the antisymmetric fields become coupled to the
vertices. We need only consider the \emph{symmetric} fields of the
interactions. This is due to the fact that we will only be interested
in the long range corrections to the background field, and the
antisymmetric fields do not produce non-analytic terms to the order at
which we are working, due to the proportionality factor of its
propagator $\sim \kappa^{2}$. In fact a diagram consisting of at least
an antisymmetric field and a graviton vertex will at least go as $\sim
\kappa^3$ which is an order higher than $\sim \kappa^{2}$. However in
a full treatment of gravitational interaction between fermionic matter
these fields will have important contributions. They will most likely
contribute to higher order calculations.

\section{The scattering matrix and the potential}\label{sec:scatt-matr-thep}
It will be fruitful to make a distinction between non-analytical and
analytical contributions from the diagrams, in order to compute the
leading long range, low energy quantum corrections to this theory.

This distinction originates from the impossibility of expanding a
massless propagator $\sim \frac{1}{q^2}$ while on the other hand we
have
\begin{equation}\label{eq:33}
\frac{1}{q^2-m^2} = -\frac{1}{m^2}\Big(1+\frac{q^2}{m^2}+\ldots \Big)
\end{equation}
expansion for the massive propagator. No $\sim \frac{1}{q^2}$ terms
are generated by the above expansion of the massive propagator. We
see that the non-analytical contributions are inherently non-local
effects which cannot be expanded in a power series in momentum. Thus
the non-analytical effects appear from the propagation of massless
particle modes, in our case the gravitons and photons. These
non-analytic contributions will be governed to leading order only by
the minimally coupled Lagrangian. 

The non-analytical effects are e.g. terms in the S-matrix which go as
$\sim \ln(-q^2)$ or $\sim\frac1{\sqrt{-q^2}}$, the general example of
an analytical effect is a power series in the momentum $q$. The
analytical contributions in the diagrams are local effects, they will
always be expandable in power series solutions. We will only consider
the non-analytical contributions here, since we are only interested in
non-local effects.

As in~\cite{Bohr1}, the high energy renormalization of the theory will
also be of no concern for us, since we are also only interested in
finding the leading finite non-analytical momentum contributions for
the 1-loop diagrams in the low energy scale of the theory. The
singular analytical momentum parts which are absorbed into
coefficients of the higher derivative couplings, will have no part to
play here, they will ultimately not be manifested in this energy
regime of the theory.
\subsubsection{Defining the Potential}\label{sec:defining-potential}
The S-matrix is defined as the scattering matrix between incoming and
outgoing particles.
The invariant matrix element $i\mathcal{M}$ originating from the diagrams is
\begin{equation}\label{eq:36}
\bk{k_{1}k_{2} \ldots}{iT}{k_{A} k_{B}} = ( 2 \pi )^{4}
\delta^{4}(k_{A}+k_{B}-\Sigma k_{\rm final})\Big( \textit{i}
\mathcal{M} \Big)
\end{equation}
here we have two incoming particles. If we Fourier transform the
earlier mentioned non-analytic terms to real space, we easily see how
the non-analytic terms contribute to the long-ranged corrections
\begin{align}
\label{eq:68} \int \frac{d^{3}q}{(2\pi)^{3}} e^{i \vec q \cdot \vec r} \frac{1}{\big|\bf
q^{2}\big|}  &= \frac{1}{4\pi r} \\ 
 \label{eq:69} \int \frac{d^{3}q}{(2\pi)^{3}} e^{i \vec q \cdot \vec r}  \frac{1}{\big|\bf
q\big|}  &= \frac{1}{2\pi^{2} r^{2}}  \\
 \label{eq:70} \int \frac{d^{3}q}{(2\pi)^{3}} e^{i \vec q \cdot \vec r} \ln{(\bf
q^{2})} &= -\frac{1}{2\pi r^{3}}
\end{align}
obviously these terms indeed do contribute to the long-range
corrections. When we calculate the tree diagrams, we explicitly see
that the non-analytic contribution of the type \eqref{eq:68}, will
correspond to the Coulomb and Newtonian part of the potentials and the
higher power of $\frac{1}{r}$ will generate the leading order and
classical corrections to the Coulomb and Newtonian
potentials. Explicitly the invariant matrix element will look like
\begin{multline}\label{eq:38}
\mathcal{M} = \Bigg(A+ B q^{2} + (\alpha_{1} \kappa^{2} +
\alpha_{2}e^{2})\frac{1}{q^{2}}\\
 + \beta_{1} \kappa^{2}e^{2}q^{2}\ln{(-q^{2})} + \beta_{2}
\kappa^{2}e^{2}q^{2}\frac{m}{\sqrt{-q^{2}}} \cdots \Bigg)
\end{multline}
where $A,B,\ldots$ correspond to the analytical, local and
short-ranged interactions, these terms will only dominate in the high
energy regime of the effective field theory, whereas
$\alpha_{1},\alpha_{2},\ldots$ and $\beta_{1},\beta_{2},\ldots$
correspond to the leading non-analytic, non-local, long-range
contributions to the amplitude. Many diagrams will yield pure analytic
contributions to the S-matrix, such diagrams will not be necessary in
our calculations, we will only consider the non-analytic contributions
from the 1-loop diagrams. The diagrams which will yield non-analytic
contributions to the S-matrix amplitude are those containing two or
more massless propagating particles.

Relating the Born approximation to the scattering amplitude in
non-relativistic quantum mechanics¨, we get in terms of $iT$
\begin{equation}\label{eq:39}
\bk{k_{1}k_{2}\ldots}{iT}{k_{A}k_{B}} = -i \widetilde{V}({\bf q}) (2\pi) \delta (E-E')
\end{equation}
where $\bf q = p'-p$ and $\widetilde{V}({\bf q})$ is the
non-relativistic potential transformed in momentum space. We should be
careful when comparing with \eqref{eq:36}, in $(i\mathcal{M})$ factors
of $(2m_{1}\times 2m_{2})$ arise due to relativistic normalization
conventions, thus we divide with these to obtain the non-relativistic
limit. Equating the two we deduce
\begin{equation}\label{eq:40}
\begin{aligned}
 -i \widetilde{V}({\bf q}) (2\pi) \delta (E-E')
&\sim  ( 2 \pi )^{4}
\delta^{4}(k_{A}+k_{B}-\Sigma k_{\rm final})\Big( \textit{i}
\mathcal{M} \Big)
\end{aligned}
\end{equation}
or rather
\begin{equation}\label{eq:41}
\tilde{V}({\bf q})=-\frac{1}{2m_{1}}\frac{1}{2m_{2}} \int
\frac{d^{3}k}{(2\pi)^{3}} (2\pi)^{3} \delta^{3}(k_{A}+k_{B}-\Sigma
k_{\rm final})(\mathcal{M})
\end{equation}
Momentum integration yields the non-relativistic potential
\begin{equation}\label{eq:42}
\tilde{V}({\bf q}) = -\frac{1}{2m_{1}}\frac{1}{2m_{2}}\mathcal{M}
\end{equation}
or in coordinate space
\begin{equation}\label{eq:43}
V({\rm x}) = -\frac{1}{2m_{1}}\frac{1}{2m_{2}} \int
\frac{d^{3}k}{(2\pi)^{3}} e^{i \bf k \cdot \bf x}  \mathcal{M}
\end{equation}
In our calculations, $\mathcal{M}$ will only contain the non-analytic
contributions of the amplitude of the scattering process to 1-loop
order, and we will not compute the full amplitude of the S-matrix,
only the long-range corrections will be of our interest. In order to
obtain their contribution to the potential, only a subclass of
scattering matrix diagrams will be required. If we wanted to find the
full total non-relativistic potential, we would merely have to include
the remaining 1-loop diagrams. This type of calculation has e.g. been
performed in~\cite{Hamber} (who also have used the same definition of
the potential as us) where the full amplitude is considered. Their
choice of potential included all 1-loop diagrams, hence they obtained
a gauge invariant definition of the potential. This choice of the
potential makes good physical sense since it is gauge invariant, but
other choices are also possible. The most convenient choice could
depend on the physical situation at hand or how the total energy is
defined. The gauge invariant choice is also equivalent to the
suggestion in~\cite{Kazakov}, where it is suggested that one should
use the full set of diagrams constituting the scattering matrix, from
which one can decide the non-relativistic potential from the total sum
of the 1-loop diagrams. However, it is worthwhile to note that we
consider all the non-analytic corrections to 1-loop order, thus if we
had the full amplitude to 1-loop order we would still need to extract
the non-analytical parts! We will continue using this definition of
the potential.
\section{Results for the Feynman diagrams}\label{sec:results-feym-diagr}
\subsection{Diagrams contributing to the non-analytic parts of the
scattering matrix potential}\label{sec:diagr-contr-non}
In this section we shall extract the non-analytical parts of a limited
set of 1-loop diagrams needed for the 1-loop scattering matrix in the
combined quantum theory of QED and general relativity (however, it is
a practically complete set of diagrams in terms of non-analytical
contributions to the scattering matrix). We will explicitly see that
the non-analytic contributions indeed correspond to the long range
corrections of the potential. This will become obvious when the
amplitudes are Fourier transformed to produce the scattering
potential, whence all the analytic pieces are disregarded. The
resulting non-analytic piece of the scattering amplitude will then be
used to construct the leading corrections to the non-relativistic
gravitational potential.
\subsection{Classical physics}\label{sec:classical-physics}
Here we will look at the tree diagrams. The fermion-fermion scattering
process at tree level should of course reproduce the results of
classical physics both for gravitational interactions as well as for
electromagnetic interactions.
\subsubsection{Tree diagrams}\label{sec:tree-diagrams} 
\begin{figure}[h]\vspace{0.5cm}
\begin{minipage}{0.43\linewidth}
\begin{center}
\includegraphics[scale=1]{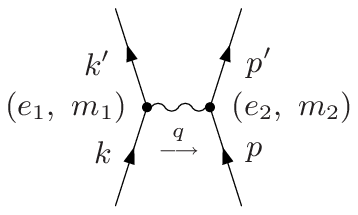}
\end{center}
{\centering 1(a)}
\end{minipage}
\begin{minipage}{0.43\linewidth}
\begin{center}
\includegraphics[scale=1]{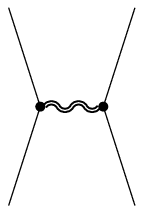}
\end{center}
{\centering 1(b)}
\end{minipage}\vspace{0.5cm}
\caption{The set of tree diagrams}\label{tree}
\end{figure}

Given in figure~\ref{tree}, we have depicted the scattering process,
where the (incoming/outgoing) momenta for particle one are $(k/k')$
with the (mass/charge) being $(m_{1}/e_{1})$, and similarly for the
second particle with the (incoming/outgoing) momenta $(p,p')$ and
$(m_{2}/e_{2})$ being the (mass/charge). This is assigned for all the
other diagrams as well. The formal expression for the digram depicted
in figure~\ref{tree}(a), the scattering process involving a photon
exchange, is
\begin{equation}\label{eq:44}
\begin{aligned}
\textit{i} \mathcal{M}_{1(a)} &= \bar u(p') [\tau^{\alpha}] u(p)\bar
u(k')[\tau^{\beta }] u(k)\Big[ - \frac{\textit{i}\eta_{\alpha \beta}}{q^{2}}  \Big]
\end{aligned}
\end{equation}
and a graviton exchange, figure~\ref{tree}(b)
\begin{equation}\label{eq:45}
\textit{i} \mathcal{M}_{1(b)} = \bar u(p') [\tau^{\mu \nu}] u(p)\bar
u(k')[\tau^{\alpha \beta }] u(k) \left[  \frac{\textit{i}\mathcal{P}_{\mu \nu \alpha \beta}}{q^{2}}  \right]
\end{equation}
yielding the well known classical results, namely the Coulomb 
\begin{equation}\label{eq:46}
\begin{aligned}
V_{1(a)}(r) &= \frac{e_{1}e_{2}}{ 4 \pi r  }
\end{aligned}
\end{equation}
and Newtonian potentials.
\begin{equation}\label{eq:47}
\begin{aligned}
V_{1(b)}(r) &= -\frac{ G m_{1}m_{2}}{ r}
\end{aligned}
\end{equation}
It is worthwhile to note that already at this stage the level of
difficulty is not obvious. There is virtually no problem in working
out the Coulomb term for the interaction, technically and
mathematically it is straightforward. However, in comparison, the
Newtonian term is much more sophisticated to work out. This is mainly
due to the many $\gamma$-matrices involved (since we are working with
fermions) and the long vertex rules theories get when coupled to
gravity. This difference will play a much bigger role when more
complicated diagrams are involved. Indeed, the next set of diagrams
were perhaps the most challenging of them all, the box and crossed box
diagrams.
\subsection{The 1PI diagrams}\label{sec:1pi-diagrams}
We will calculate all the relevant 1PI diagrams necessary to find the
long range corrections to the potentials. We will start with the box
and crossed box diagrams and continue with the set of triangular
diagrams. Lastly we will work out the circular loop diagram.
\subsubsection{The box and crossed box diagrams}\label{sec:box-crossed-box}
\begin{figure}[h]\vspace{0.5cm}
\begin{minipage}{0.43\linewidth}
\begin{center}
\includegraphics[scale=1]{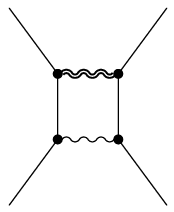}
\end{center}
{\centering 2(a)}
\end{minipage}
\begin{minipage}{0.43\linewidth}
\begin{center}
\includegraphics[scale=1]{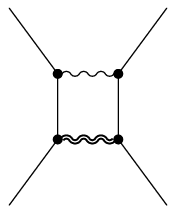}
\end{center}
{\centering 2(b)}
\end{minipage}\vspace{0.5cm}
\begin{minipage}{0.43\linewidth}
\begin{center}
\includegraphics[scale=1]{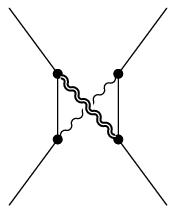}
\end{center}
{\centering 2(c)}
\end{minipage}
\begin{minipage}{0.43\linewidth}
\begin{center}
\includegraphics[scale=1]{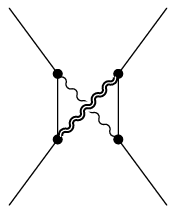}
\end{center}{\centering 2(d)}
\end{minipage}
\caption{The set of box diagrams contributing to the potential}\label{box}
\end{figure}

There are in all four distinct diagrams. Two box and two crossed box
diagrams, these are depicted in figure~\ref{box}.
We will not treat all of these diagrams here. We will only look at one
of these, the rest can be treated in the same manner. Explicitly
diagram \ref{box}(a) is defined by
\begin{equation}\label{eq:48}
\begin{aligned}
\textit{i} \mathcal{M}_{2(a)} &= \int \frac{d^{4}\ell}{(2 \pi)^{4}}  \Big[- \frac{\textit{i} \eta_{\delta \gamma}}{\ell^{2}} \Big] \Big[
\frac{\textit{i} \mathcal{P}_{\mu \nu \rho \sigma}}{( \ell + q )^{2}}
\Big]\\
&\times \bar u (p') \Big[ \tau^{\rho \sigma }( p - \ell , p' ) D_F( p
- \ell )
\tau^{\delta} ( p , p - \ell ) \Big] u(p)\\
&\times \bar u (k') \Big[ \tau^{\mu \nu }( k + \ell , k' ) D_{F}( k +
\ell )
\tau^{\gamma} ( k , k + \ell ) \Big] u(k)
\end{aligned}
\end{equation}

The methods and techniques to work these diagrams are identical to
those shown in~\cite{Bohr1}, we will repeat them shortly here.

The only difference lies in the fact that these diagrams require four
different integrals that were not worked out previously. We have
worked them out, and the coefficients can be obtained by contacting
us, they are too tedious to be written down in the appendix. Other
than these integrals, these diagrams simply had to be worked out even
though they involved enormous amounts of calculations. The level of
difficulty is much larger than in the previous case, due to the
reasons mentioned earlier. All the box diagrams have been calculated
by using symbolic manipulation on a computer. These have partly been
checked by hand.
\begin{widetext}
On the mass shell we will have the following type of identities
\begin{gather}\label{eq:49}
\ell \cdot q = \frac{1}{2} ( (\ell +q)^{2} - q^{2} -\ell^{2}  ), \qquad 
\ell \cdot k = \frac{1}{2} ( (\ell +k)^{2} - m_{1}^{2} -\ell^{2}  ),\qquad
\ell \cdot p = -\frac{1}{2} ( (\ell -p)^{2} - m_{2}^{2} -\ell^{2}  )
\end{gather}
so
\begin{multline}\label{eq:50}
q_{\mu}K^{\mu\nu } = \int \frac{d^{4}\ell}{(2 \pi)^{4}} \frac{(\ell
\cdot q)\ell^{\nu}}{\ell^{2} (\ell + q)^{2} [ ( \ell + k )^{2} -
m_{1}^{2} ] [ ( \ell - p )^{2} - m_{2}^{2} ]}\\
\rightarrow - \frac{q^{2}}{2} \int \frac{d^{4}\ell}{(2 \pi)^{4}}
\frac{\ell^{\nu }}{\ell^{2} (\ell + q)^{2} [ ( \ell + k )^{2} -
m_{1}^{2} ] [ ( \ell - p )^{2} - m_{2}^{2} ]} =-
\frac{q^{2}}{2}K^{\nu}
\end{multline}
since the terms with $(\ell+q)^{2}$ and $\ell^{2}$ simply do not
contribute with non-analytical results. 

A more important reduction of the integrals is with contraction of
the sources momenta rather than the exchange momentum
\begin{align}\label{eq:51}
k_{\mu}K^{\mu\nu } &= \int \frac{d^{4}\ell}{(2 \pi)^{4}} \frac{(\ell
\cdot k)\ell^{\nu}}{\ell^{2} (\ell + q)^{2} [ ( \ell + k )^{2} -
m_{1}^{2} ] [ ( \ell - p )^{2} -
m_{2}^{2} ]}\rightarrow \frac{1}{2}\int \frac{d^{4}\ell}{(2 \pi)^{4}} \frac{ \ell^{\nu}
}{\ell^{2} (\ell + q)^{2} [ ( \ell - p )^{2} - m_{2}^{2} ]}=\frac{1}{2}I_{p}^{\nu}
\end{align}
or in similar manner
\begin{align}\label{eq:52}
p_{\mu}K^{\mu\nu } &= \int \frac{d^{4}\ell}{(2 \pi)^{4}} \frac{(\ell
\cdot p)\ell^{\nu}}{\ell^{2} (\ell + q)^{2} [ ( \ell + k )^{2} -
m_{1}^{2} ] [ ( \ell - p )^{2} - m_{2}^{2} ]}\rightarrow-
\frac{1}{2}\int \frac{d^{4}\ell}{(2 \pi)^{4}} \frac{ \ell^{\nu}
}{\ell^{2} (\ell + q)^{2} [ ( \ell + k )^{2} - m_{1}^{2}
]}=-\frac{1}{2}I_{k}^{\nu}
\end{align}
where the subscripts $_{k}$ and $_{p}$ on the $I$'s are written to
indicate that the propagators left in the integrals are either from
the particle with momentum $k$ or $p$. Thus a loop momentum contracted
with a source momentum simplifies our integrals considerably.
\end{widetext}
The kinematics are (on the mass shell)
\begin{gather}\label{eq:53}
k  \cdot q =
p' \cdot q = \frac{q^{2}}{2}  \qquad
k' \cdot q =
p  \cdot q = -\frac{q^{2}}{2}  \\
k \cdot k' = m_{1}^2 - \frac{q^{2}}{2}  \qquad
p \cdot p' = m_{2}^2 - \frac{q^{2}}{2} 
\end{gather}
The potential contribution from these diagrams are found to be
\begin{multline}\label{eq:54}
V(r)_{2(a)+2(b)+2(c)+2(d)} = \frac{3}{4} \frac{e_{1}
e_{2}(m_{1}+m_{2})G}{\pi r^{2}} \\
- \frac{118}{48} \frac{ G e_{1} e_{2}}{ \pi^{2}r^{3}}
\end{multline}
These diagrams yield both a classical contribution - the $\sim
\frac{1}{r^{2}}$ - and a quantum correction - the $\sim
\frac{1}{r^{3}}$. In the scalar QED calculations the box diagrams only
generated quantum corrections, it is interesting to see that the
Feynman diagrams do not necessarily generate identical results for the
individual diagrams. This will become more clear in the other
diagrams.
\subsubsection{The triangular diagrams}\label{sec:triangular-diagrams}
\begin{figure}[h]
\begin{minipage}{0.43\linewidth}
\begin{center}
\includegraphics[scale=1]{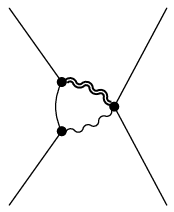}
\end{center}
{3(a)}
\end{minipage}
\begin{minipage}{0.43\linewidth}\vspace{0.4cm}
\begin{center}
\includegraphics[scale=1]{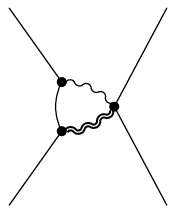}
\end{center}
{\vspace{0.15cm}3(b)}
\end{minipage}\vspace{0.3cm}
\begin{minipage}{0.43\linewidth}
\begin{center}
\includegraphics[scale=1]{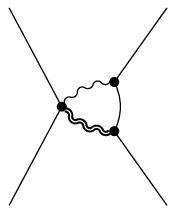}
\end{center}
{3(c)}
\end{minipage}
\begin{minipage}{0.43\linewidth}
\begin{center}
\includegraphics[scale=1]{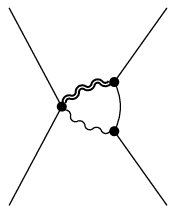}
\end{center}
{3(d)}
\end{minipage}
\caption{The set of triangular diagrams contributing non-analytically
to the potential.}\label{tri}
\end{figure}

Diagrammatically it is given in figure~\ref{tri}, these are the only
triangular diagrams that contribute with non-analytic terms.
We will again only consider one instance of these diagrams, namely the
first fig.~\ref{tri}(a). Formally it is written down as follows
\begin{multline}\label{eq:55}
\textit{i} \mathcal{M}_{3(a)} = \int \frac{d^{4}\ell}{(2 \pi)^{4}}
\bar u (p') \Big[ \tau^{ \rho \sigma (\delta)} \Big] u(p)\\
\times \bar u (k') \Big[ \tau^{\mu \nu }( -\ell - k , k' ) D_{F}(
-\ell - k ) \tau^{\gamma} ( k , -\ell - k ) \Big] u(k)\\
\times \Big[- \frac{\textit{i} \eta_{\delta \gamma}}{\ell^{2}} \Big] \Big[
\frac{\textit{i} \mathcal{P}_{\mu \nu \rho \sigma}}{( \ell + q )^{2}}
\Big]
\end{multline}
Upon applying contractions and insertion of the relevant integrals
where after Fourier transformations are performed, we end up with the
potential contribution
\begin{equation}\label{eq:56}
\begin{aligned}
V_{3(a)+3(b)+3(c)+3(d)}(r) & =- \frac{9 G e_{1} e_{2}}{4 \pi^{2}r^{3}}
\end{aligned}
\end{equation}
which is also different from the result obtained in the~\cite{Bohr1}
for the triangular diagrams.
\subsubsection{The circular diagram}\label{sec:circular-diagram}
\begin{figure}[h!]
\begin{center}
\includegraphics[scale=1]{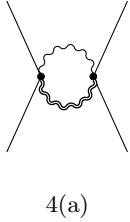}
\end{center}
{4(a)}
\caption{The circular diagram with non-analytic contributions.}\label{circ}
\end{figure}

The circular diagram is depicted in figure \ref{circ}.
Formally it can be written as
\begin{multline}\label{eq:57}
\textit{i} \mathcal{M}_{4(a)} = \int \frac{d^{4}\ell}{(2 \pi)^{4}} \bar u
(p') \Big[ \tau^{ \rho \sigma (\delta)} \Big] u(p) \Big[-
\frac{\textit{i} \eta_{\delta \gamma}}{\ell^{2}} \Big]\\
\times \Big[ \frac{\textit{i} \mathcal{P}_{\mu \nu \rho \sigma}}{(
\ell + q )^{2}} \Big] \bar u (k') \Big[ \tau^{ \mu \nu (\gamma)} \Big]
u(k)
\end{multline}
When doing all the contractions and rearranging the $\gamma$-matrices one
arrives at the result that the contribution to the potential from the
circular loop diagram is precisely equal to nill.
\begin{equation}\label{eq:58}
 V_{4(a)}=0
\end{equation}
Which indeed is very different from the result obtained in~\cite{Bohr1}!
\subsection{The vertex correction diagrams}\label{sec:vert-corr-diagr-1}
There are several sets of 1PR diagrams. All these are presented in
this section.
\subsubsection{The 1PR diagram}\label{sec:1pr-diagram}
\begin{figure}[h]\vspace{0.5cm}
\begin{minipage}{0.43\linewidth}
\begin{center}
\includegraphics[scale=1]{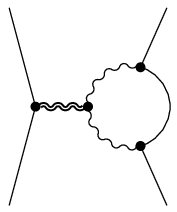}
\end{center}
{\centering 5(a)}
\end{minipage}
\begin{minipage}{0.43\linewidth}
\begin{center}
\includegraphics[scale=1]{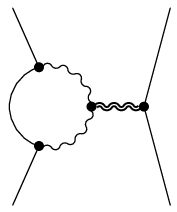}
\end{center}
{\centering 5(b)}
\end{minipage}\vspace{0.5cm}
\caption{The first set of 1PR diagrams contributing non-analytically to the potential}\label{1PR-1}
\end{figure}

The first set of these 1PR diagrams is given in figure~\ref{1PR-1}.
These diagrams are the only ones corresponding to the gravitational
vertex corrections. Again we will only consider one instance of these
diagrams. The matrix element originating from figure~\ref{1PR-1}(a)
diagram is
\begin{multline}\label{eq:59}
 \textit{i}\mathcal{M}_{5(a)} = \int
\frac{d^4\ell}{(2\pi)^4}\bar u(p') \Big[ \tau^\beta D_F(p-\ell)
\tau^\alpha \\
\times  \tau^{\rho\sigma(\gamma\delta)}(\ell,\ell+q) \Big] u(p)
\bigg[ \frac{-i \eta_{\alpha\gamma}}{\ell^2} \bigg] \bigg[ \frac{-i
\eta_{\beta\delta}}{(\ell+q)^2} \bigg]\\
\times \bigg[ \frac{i {\cal P}_{\mu\nu\rho\sigma}}{q^2} \bigg] \bar u
( k' )[ \tau^{\mu\nu}(k,k')] u(k)
\end{multline}
which yields the following potential when all the diagrams are summed
\begin{equation}\label{eq:60}
\begin{aligned}
  V_{5(a)+5(b)}(r) & = G \left(\frac{m_{2} e_{1}^{2} + m_{1}
e_{2}^{2}}{8 \pi r^{2}}-\frac{
e_{1}^{2}\frac{m_{2}}{m_{1}}+e_{2}^{2}\frac{m_{1}}{m_{2}} }{ 3 \pi^{2}
r^{3} } \right)
\end{aligned}
\end{equation}
This checks with~\cite{Donoghue} where it has also been calculated.\\

\begin{figure}[h]\vspace{0.5cm}
\begin{minipage}{0.43\linewidth}
\begin{center}
\includegraphics[scale=1]{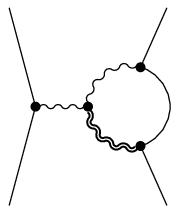}
\end{center}
{\centering 6(a)}
\end{minipage}
\begin{minipage}{0.43\linewidth}
\begin{center}
\includegraphics[scale=1]{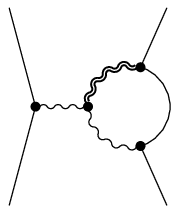}
\end{center}
{\centering 6(b)}
\end{minipage}\vspace{0.5cm}
\begin{minipage}{0.43\linewidth}
\begin{center}
\includegraphics[scale=1]{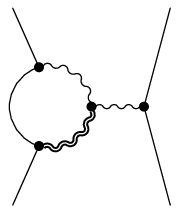}
\end{center}
{\centering 6(c)}
\end{minipage}
\begin{minipage}{0.43\linewidth}
\begin{center}
\includegraphics[scale=1]{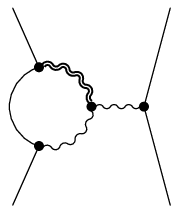}
\end{center}
{\centering 6(d)}
\end{minipage}
\caption{The second set of 1PR diagrams contributing to the potential}\label{1PR-2}
\end{figure}

Of the next set of the 1PR diagrams, depicted in figure~\ref{1PR-2} we
will also only consider the first one.
This is the first of the set of diagrams corresponding to the photonic
vertex corrections. Formally diagram~\ref{1PR-2}(a) is given by
\begin{multline}\label{eq:61}
 \textit{i}\mathcal{M}_{6(a)} = \int \frac{d^4\ell}{(2\pi)^4}\bar u(p')
\Big[ \tau^\beta D_F(p-\ell) \tau^{ \mu \nu } ( p , p - \ell ) \Big] u(p)\\
\times\bigg[ \frac{-i \eta_{\alpha\beta}}{ ( \ell + q )^2 } \bigg] \bigg[
\frac{ i \mathcal{ P }_{ \rho \sigma \mu \nu }}{ \ell^2} \bigg] [
\tau^{\rho\sigma( \delta \alpha ) }( q , \ell + q ) ]\\
 \times \bigg[ \frac{-i \eta_{\gamma\delta}}{ q^2 } \bigg] \bar u(k')[\tau^{\gamma}] u(k)
\end{multline}
giving the potential
\begin{equation}\label{eq:62}
\begin{aligned}
V_{6(a)+6(b)+6(c)+6(d)}(r)  & =- \frac{3 G e_{1} e_{2}}{4 \pi^{2}r^{3}}
\end{aligned}
\end{equation}\\

\begin{figure}[h]\vspace{0.5cm}
\begin{minipage}{0.43\linewidth}
\begin{center}
\includegraphics[scale=1]{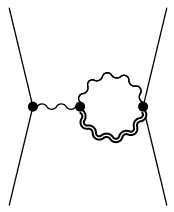}
\end{center}
{\centering 7(a)}
\end{minipage}
\begin{minipage}{0.43\linewidth}
\begin{center}
\includegraphics[scale=1]{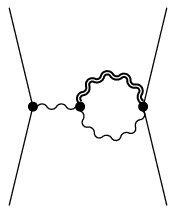}
\end{center}
{\centering 7(b)}
\end{minipage}\vspace{0.5cm}
\begin{minipage}{0.43\linewidth}
\begin{center}
\includegraphics[scale=1]{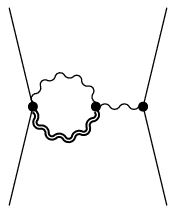}
\end{center}
{\centering 7(c)}
\end{minipage}
\begin{minipage}{0.43\linewidth}
\begin{center}
\includegraphics[scale=1]{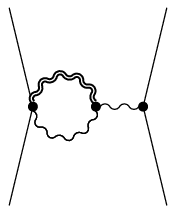}
\end{center}
{\centering 7(d)}
\end{minipage}
\caption{The third set of 1PR diagrams contributing to the potential}\label{1PR-3}
\end{figure}

The second set of diagrams corresponding to the photonic vertex
corrections are given in figure~\ref{1PR-3}. Formally diagram \ref{1PR-3}(a)
is given by
\begin{equation}\label{eq:63}
\begin{aligned}
i \mathcal{M}_{7(a)} = &\int \frac{ d^4 \ell }{( 2 \pi )^4 } \tau^{
\mu \nu ( \delta \alpha ) } ( q, q + \ell ) \Big[ \frac{ i \mathcal{ P
}_{ \mu \nu \sigma \rho } }{ \ell^2 }  \Big] \Big[ \frac{ - i \eta_{
\alpha \beta } }{ ( \ell + q )^2 }  \Big]\\
 &\times \bar u( p' )[ \tau^{ \sigma
\rho ( \beta ) } ( p , p', e_{2} )] u( p ) \Big[ \frac{ - i \eta_{
\gamma\delta } }{ q ^2 }  \Big]\\
&\times \bar u (k')[\tau^{\gamma}] u(k)
\end{aligned}
\end{equation}
yielding the potential
\begin{equation}\label{eq:64}
\begin{aligned}
V_{7(a)+7(b)+7(c)+7(d)}(r) & = \frac{3 G e_{1} e_{2}}{2 \pi^{2}r^{3}}
\end{aligned}
\end{equation}
\subsubsection{The vacuum polarization diagram}\label{sec:vacu-polar-diagr}
\begin{figure}[h]\vspace{0.5cm}
\begin{center}
\includegraphics[scale=1]{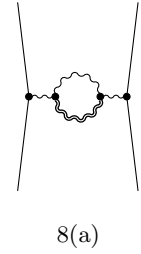}
\end{center}
{8(a)}
\caption{The vacuum polarization diagram contribution to the non-relativistic potential.}\label{vac}
\end{figure}

The diagram is depicted in figure~\ref{vac}. The formal expression
for this diagram is
\begin{equation}\label{eq:65}
\begin{aligned}
  i \mathcal{M}_{8(a)} = &\int \frac{ d^4 \ell }{( 2 \pi )^4 } \bar
u(p') [\tau^{\gamma}] u(p) \bigg[\frac{ -i \eta_{ \gamma\delta } }{
q^2 } \bigg]\\
&\times \tau^{ \sigma \rho ( \delta \alpha ) } ( q, -\ell )
\bigg[\frac{ -i \eta_{ \alpha \beta } }{ \ell^2 } \bigg] \bigg[ \frac{
i \mathcal{ P }_{ \mu \nu \rho \sigma }}{ ( \ell + q )^2 } \bigg]\\
&\times \tau^{ \mu \nu ( \beta \epsilon ) } ( -\ell , q ) \bigg[\frac{
-i \eta_{ \phi\epsilon } }{ q^2 } \bigg] \bar u(k') [\tau^{\phi}] u(k)
\end{aligned}
\end{equation}
This is the only instance in these calculations that $\gamma$-matrices
are not explicitly involved. Simple index contractions are done and one obtains
the potential contribution after going to the non relativistic limit
\begin{equation}\label{eq:66}
V_{8(a)}(r) = \frac{G e_{1} e_{2}}{6 \pi^{2}r^{3}}
\end{equation}
which is identical with the bosonic version.
\section{The Results for the Potential}\label{sec:results-potential}
When adding up all the separate contributions, we end up with
\begin{equation}\label{eq:67}
\begin{aligned}
V(r)= &-\frac{ G m_{1}m_{2}}{ r} +
\frac{\tilde{\alpha}\tilde{e}_{1}\tilde{e}_{2}}{ r }\\
& +\frac{1}{2} \Big(m_{2} \tilde{e}_{1}^{2} + m_{1} \tilde{e}_{2}^{2}
\Big) \frac{G \tilde{\alpha}}{c^{2}r^{2}}+
3\frac{\tilde{e_{1}}\tilde{e_{2}}(m_{1}+m_{2})\tilde{\alpha}G}{c^{2}r^{2}}\\
 &- \frac{4}{3} \frac{\tilde{G \alpha} \hbar}{\pi c^{3}r^{3}} \Big(
e_{1}^{2}\frac{m_{2}}{m_{1}}+e_{2}^{2} \frac{m_{1}}{m_{2}}\Big)
-15{\mbox{$\frac{1}{6}$}} \frac{ G \tilde{\alpha}\hbar
\tilde{e}_{1} \tilde{e}_{2}}{ \pi r^{3}}
\end{aligned}
\end{equation}
where we have included the appropriate physical factors of $\hbar$, $c$
and we have further rescaled everything in terms of
$\tilde{\alpha}=\frac{\hbar c}{137}$, lastly
$(\tilde{e}_{1},\tilde{e}_{2})$ are the normalized charges in units of
elementary charge. The result is divided into three separate parts,
the first two terms represent the Newtonian and Coulomb potentials,
the next two terms represent the classical post-Newtonian corrections
to the potential, which also can be found by pure classical treatment
of general relativity with the inclusion of charged matter
sources~\cite{Barker}. It is interesting to see that loop calculations
also reproduce classical results, and not only quantum
corrections. Finally, the last two terms are the leading 1-loop
quantum corrections. We have in a greater extent reproduced the
results of~\cite{Bohr1}, except for the last quantum correction where
we get the factor $(-15\mbox{$\frac{1}{6}$})$ instead of $(-8)$ as
in~\cite{Bohr1}.

It is interesting to note how the different terms come about in the
corrections to the potential. The classical terms, the Newtonian and
Coulomb terms, of course originate from the tree diagrams alone. The
first post-Newtonian term and the first quantum correction notably
originate from one type of diagrams alone, namely the gravitational
vertex correction, see figure~\ref{1PR-1}. These match~\cite{Bohr1}
exactly. Now the second post-Newtonian correction originates from all
the box diagrams and nothing else, which indeed also is in full
agreement with~\cite{Bohr1}. However, the second quantum correction is
a sum of partly the box contributions and the rest of the
diagrams. This correction does not match the quantum correction in
ref.~\cite{Bohr1}. The rest of the diagrams have been done by hand,
and all the diagrams done in this paper have been checked by symbolic
manipulation on a computer.

\section{Discussion}\label{sec:discussion}
We have examined the leading order quantum corrections to
gravitational coupling of a spin-{\mbox{\small$\frac{1}{2}$}} massive
charged particle. Explicitly we have extracted the non-analytic terms
from the diagrams, which exactly manifested themselves as corrections
to the long range forces, this was realized when we Fourier
transformed these terms into coordinate space. These terms originated
from the propagation of the massless particles, here the photons and
gravitons. We have obtained similar results for most of the
contributions to the corrections of the potentials, when compared
with~\cite{Bohr1}. Only one of the leading quantum corrections differs
from the bosonic calculation.

One could in similar manner do a QED - pure gravity scattering
calculation as in~\cite{Holstein}. However, more obstacles would be
needed to overcome. First of all, one would have to find many new
vertex rules involving the antisymmetric fields. Moreover we would
have to derive second order Lagrangian densities both in terms of the
symmetric and antisymmetric fields. Indeed, in this direction there
lies a considerably interesting project ahead.

\begin{acknowledgments}\label{sec:discussion-1}
I would like to thank Poul~Henrik~Damgaard for his supervision,
guidance and for many suggestions. I would also like to thank
N.~E.~J.~Bjerrum-Bohr for invaluable discussions, and also for
supervision and guidance during the initial stage of this work.
\end{acknowledgments}

\appendix

\section{Feynman rules}\label{sec:feynman-rules}

\subsection{Propagators}\label{sec:propagators}
The relevant propagators are presented in this section.

\subsubsection{Photon propagator}\label{sec:photon-propagator}
The photon propagator is no stranger in QFT. In Feynman gauge the
propagator becomes
\begin{align*}
  \parbox{25mm}{\includegraphics[scale=1]{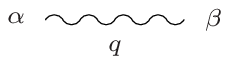}
        }   &= \quad \frac{-i \eta_{\alpha\beta}}{q^2+i\epsilon}
\end{align*}
\subsubsection{Graviton propagator}
The graviton propagator is perhaps a stranger. However it has been
worked out several places. In the Harmonic gauge we get the following
for the graviton propagator
\begin{align*}
 \parbox{25mm}{ \includegraphics[scale=1]{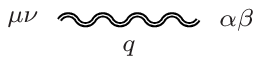} }   &= \quad \frac{i \mathcal{P}_{\mu\nu\alpha\beta}}{q^2+i\epsilon}
\end{align*}
with the projection operator
\begin{align*}
 \mathcal{P}_{\mu\nu\alpha\beta} &= \frac{\frac{1}{2} (\eta_{\alpha\{\mu}\eta_{\nu\}\beta}-\eta_{\mu\nu}\eta_{\alpha\beta})}{q^2+i\epsilon} 
\end{align*}

\subsubsection{Fermion propagator}
The fermion propagator can be found many places in literature, it is
very well known
\begin{align*}
  \parbox{20mm}{ \includegraphics[scale=1]{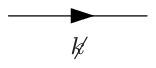}       }   &= \quad \frac{i}{( \slash k - m )} \quad = \quad \frac{i( \slash k + m )}{ k^2 - m^2 }
\end{align*}

\subsection{Vertices}
The vertices are presented here. They are all derived
in~\cite{Thesis}. For all vertices the rules of momentum conservation
has been applied.

\subsubsection{1-photon-2-fermion vertex}
The 1-photon-2-fermion vertex can also be looked up in literature
it is worked out to be\\
\begin{minipage}[h]{.5\textwidth}
\begin{equation*}
  \parbox{25mm}{\includegraphics[scale=1]{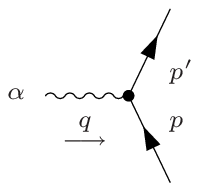} } = \quad
\tau^{\alpha} (p,p')\phantom{^{\beta(\gamma\delta)}}
\end{equation*}
\end{minipage}
with
\begin{align*}
 \tau^{\alpha} (p,p')= \textit{i} e_{q} \gamma^{\alpha }
\end{align*}

\subsubsection{1-graviton-2-fermion vertex}
The 1-graviton-2-fermion vertex is found to be\\
\begin{minipage}[h]{.5\textwidth}
\begin{equation*}
 \parbox{25mm}{\includegraphics[scale=1]{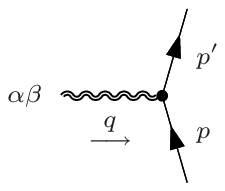}} = \quad \tau^{\alpha\beta} (p,p')\phantom{^{(\gamma\delta)}}
\end{equation*}
\end{minipage}
where
\begin{multline*}
 \tau^{\alpha\beta} (p,p') \quad = \quad \frac{ i \kappa }{ 2 } \Big[
\eta^{ \alpha \beta } \big( \frac12 ( \slash p + \slash p' ) - m \big)\\
- \frac14 \gamma^{\{\alpha} ( p + p' )^{\beta\}} \Big]
\end{multline*}

\subsubsection{1-photon-1-graviton-2-fermion vertex}
The 1-photon-1-graviton-2-fermion vertex is not known
from literature, however it is found to be\\
\begin{minipage}[h]{.5\textwidth}
\begin{equation*}
 \parbox{25mm}{\includegraphics[scale=1]{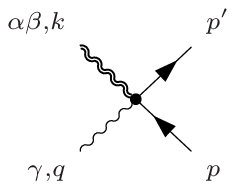}} = \quad
\tau^{\alpha\beta(\gamma)} (p,p') \phantom{^{\delta}}
\end{equation*}
\end{minipage}\\
\phantom{}\\
where
\begin{align*}
 \tau^{\alpha\beta(\gamma)} (p,p') \quad = \quad \frac{ i \kappa e_{q}}{ 4
} \gamma_a ( 2 \eta^{\gamma a} \eta^{ \alpha \beta } - \eta^{ \gamma
\{ \alpha } \eta^{ \beta \} a } )
\end{align*}

\subsubsection{1-graviton-2-photon vertex}
We have derived the following for the 1-graviton-2-photon vertex\\
\begin{minipage}[h]{.5\textwidth}
\begin{equation*}
\parbox{25mm}{
\includegraphics[scale=1]{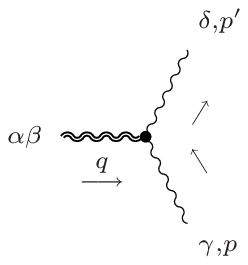}} = \quad
\tau^{\alpha\beta(\gamma\delta)} (p,p')
\end{equation*}
\end{minipage}
where
\begin{multline*}
\tau^{\alpha\beta(\gamma\delta)} (p,p') = i\kappa \bigg[ \mathcal{P}^{
\alpha \beta ( \gamma \delta ) } (p \cdot p') + \frac12 \big( \eta^{
\alpha\beta } p^\delta p'^\gamma\\
+ \eta^{ \gamma \delta } p^{ \{ \beta } p'^{ \alpha \} } - p^{ \delta
} p'^{ \{ \alpha} \eta^{ \beta \} \gamma } - p'^{ \gamma } p^{ \{
\alpha} \eta^{ \beta \} \delta } \big)\bigg]
\end{multline*}
$\mathcal{P}^{ \alpha \beta ( \gamma \delta )}$ is defined as above
and $\{ \}$ do not involve any symmetrization factors, they merely
exchange the indices.

\onecolumngrid

\section{Table of relevant Integrals} 
The following integrals are needed, note that in these integrals only
the lowest order non-analytical terms are presented
\begin{eqnarray}\displaystyle
J=&\displaystyle \int\frac{d^4\ell}{(2\pi)^4} \frac{1}{\ell^2(\ell+q)^2} & = \frac{i}{32\pi^2}\big[-2L\big] + \ldots\\
J_\mu=&\displaystyle\int\frac{d^4\ell}{(2\pi)^4} \frac{\ell_\mu}{\ell^2(\ell+q)^2} & =   \frac{i}{32\pi^2}\Big[q_\mu L\Big]+\ldots\\
J_{\mu\nu}=&\displaystyle\int\frac{d^4\ell}{(2\pi)^4} \frac{\ell_\mu \ell_\nu}{\ell^2(\ell+q)^2} 
& = \frac{i}{32\pi^2}\bigg[q_\mu q_\nu \Big(-\frac23L\Big) -
q^2\eta_{\mu\nu}\Big(-\frac16 L\Big)\bigg]+\ldots 
\end{eqnarray}
together with
\begin{eqnarray}\displaystyle
I&=&\displaystyle\int\frac{d^4\ell}{(2\pi)^4} \frac{1}{\ell^2(\ell+q)^2((\ell+k)^2-m^2)}
  = \frac{i}{32\pi^2m^2}\big[-L-S\big]+\ldots\\
I_\mu&=&\displaystyle\int\frac{d^4\ell}{(2\pi)^4}
\frac{\ell_\mu}{\ell^2(\ell+q)^2((\ell+k)^2-m^2)}\nonumber\\
 & = & \frac{i}{32\pi^2m^2}\bigg[k_\mu\bigg(\Big(-1-\frac12 \frac{q^2}{m^2}\Big)L-
\frac14\frac{q^2}{m^2}S\bigg)+q_\mu\bigg(L+\frac12S\bigg)\bigg]+\ldots\\
I_{\mu\nu}&=&\displaystyle\int\frac{d^4\ell}{(2\pi)^4} \frac{\ell_\mu \ell_\nu}{\ell^2(\ell+q)^2((\ell+k)^2-m^2)}\nonumber\\
& =&  \frac{i}{32\pi^2m^2}\bigg[q_\mu q_\nu\bigg(-L-\frac38 S\bigg)
+k_\mu k_\nu \bigg(-\frac12\frac{q^2}{m^2}L-\frac18\frac{q^2}{m^2}S\bigg)\nonumber\\
&&+\big(q_\mu k_\nu + q_\nu k_\mu \big)\bigg(\Big(\frac12 + \frac12\frac{q^2}{m^2}\Big)L + \frac{3}{16}\frac{q^2}{m^2}S\bigg)+
q^2\eta_{\mu\nu}\Big(\frac14L+\frac18S\Big)
\bigg]+\ldots
\end{eqnarray}
\begin{eqnarray}\displaystyle
I_{\mu\nu\alpha}&=&\displaystyle\int\frac{d^4\ell}{(2\pi)^4}
\frac{\ell_\mu \ell_\nu \ell_\alpha}{\ell^2(\ell+q)^2((\ell+k)^2-m^2)}\nonumber\\
& =& \frac{i}{32\pi^2m^2}\bigg[ q_\mu q_\nu
q_\alpha\bigg(L+\frac5{16}S\bigg)+k_\mu k_\nu k_\alpha\bigg(-\frac16
\frac{q^2}{m^2}L\bigg) \nonumber\\
 \nonumber&&+\big(q_\mu k_\nu k_\alpha + q_\nu k_\mu k_\alpha +
q_\alpha k_\mu k_\nu\big)\bigg(\frac13\frac{q^2}{m^2}L+
\frac1{16}\frac{q^2}{m^2}S\bigg)\nonumber \\
&&+\big(q_\mu q_\nu k_\alpha + q_\mu q_\alpha k_\nu + q_\nu q_\alpha
k_\mu \big)\bigg(\Big(-\frac13 -
\frac12\frac{q^2}{m^2}\Big)L -\frac{5}{32}\frac{q^2}{m^2}S\bigg)\nonumber \\
\nonumber &&+\big(\eta_{\mu\nu}k_\alpha + \eta_{\mu\alpha}k_\nu +
\eta_{\nu\alpha}k_\mu\big)\Big(\frac1{12}q^2L\Big) \\
&&+\big(\eta_{\mu\nu}q_\alpha + \eta_{\mu\alpha}q_\nu +
\eta_{\nu\alpha}q_\mu\big)\Big(-\frac16q^2L -\frac1{16}q^2S\Big)
\bigg]+\ldots
\end{eqnarray}
where $S=\frac{\pi^2m}{\sqrt{-q^2}}$ and $L=\ln(-q^2)$. The ellipses
denote higher order non-analytical contributions as well as the
neglected analytical terms. Please note that there seems to be a typo
in~\cite{Bohr1}, in $I_{\mu\nu\alpha}$ the factor after $(k_\mu k_\nu
k_\alpha)$ is lacking an $L$. Other than this typo all the integrals
have been checked and are agreed upon.

The following integrals are needed to do the box diagrams.
\begin{eqnarray}\displaystyle
K & = &\displaystyle \int\frac{d^4\ell}{(2\pi)^4} \frac{1}{\ell^2(\ell+q)^2((\ell+k)^2-m_1^2)((\ell-p)^2-m_2^2)}
   = \displaystyle \frac{i}{16\pi^2m_1 m_2 q^2}\bigg[\Big(1-\frac{w}{3m_1m_2}\Big)L\bigg]+\ldots \\
K'& = &\displaystyle \int\frac{d^4\ell}{(2\pi)^4} \frac{1}{\ell^2(\ell+q)^2((\ell+k)^2-m_1^2)((\ell+p')^2-m_2^2)}
   = \displaystyle  \frac{i}{16\pi^2m_1 m_2 q^2}\bigg[\Big(-1+\frac{W}{3m_1m_2}\Big)L\bigg]+\ldots\\
K^{\mu}& = &\displaystyle \int \frac{d^{4}\ell}{(2 \pi)^{4}} \frac{\ell^{\mu}}{\ell^{2}
(\ell + q)^{2} [ ( \ell + k )^{2} - m_{1}^{2} ] [ ( \ell - p )^{2} - m_{2}^{2} ]}
 =  \alpha q^{\mu} + \beta k^{\mu} + \gamma p^{\mu}\\
K'^{\mu}& = &\displaystyle \int \frac{d^{4}\ell}{(2 \pi)^{4}}
\frac{\ell^{\mu}}{\ell^{2} (\ell + q)^{2} [ ( \ell + k )^{2} -
m_{1}^{2} ] [ ( \ell + p' )^{2} - m_{2}^{2} ]}
 =  \alpha' q^{\mu} + \beta' k^{\mu} + \gamma' p'^{\mu}\\
K^{\mu\nu }& = &\displaystyle \int
\frac{d^{4}\ell}{(2 \pi)^{4}} \frac{\ell^{\mu}\ell^{\nu}}{\ell^{2}
(\ell + q)^{2} [ ( \ell + k )^{2} - m_{1}^{2} ] [ ( \ell - p )^{2} -
m_{2}^{2} ]}\nonumber\\
 &=& \Big[ q_{\mu} q_{\nu}a + k_{\mu} k_{\nu}b +
p_{\mu} p_{\nu}c\Big] +\Big[( q_{\mu} k_{\nu}+ q_{\nu} k_{\mu})d +( q_{\mu} p_{\nu}+ q_{\nu} p_{\mu})e
\Big] +( p_{ \mu} k_{\nu }+ p_{ \nu} k_{\mu })f + \eta_{\mu \nu }q^{2}g\\
K'^{\mu\nu }& = &\displaystyle \int \frac{d^{4}\ell}{(2 \pi)^{4}}
\frac{\ell^{\mu}\ell^{\nu}}{\ell^{2} (\ell + q)^{2} [ ( \ell + k )^{2}
- m_{1}^{2} ] [ ( \ell + p' )^{2} - m_{2}^{2} ]}\nonumber\\
 &=& \Big[ q_{\mu} q_{\nu}a' + k_{\mu} k_{\nu}b' + p'_{\mu}
p'_{\nu}c'\Big] +( q_{\mu} k_{\nu}+ q_{\nu} k_{\mu})d' +( q_{\mu}
p'_{\nu}+ q_{\nu} p'_{\mu})e' +( p'_{ \mu} k_{\nu }+ p'_{ \nu} k_{\mu })f'
+ \eta_{\mu \nu }q^{2}g'
\end{eqnarray}
Here we have defined $w = (k\cdot p)-m_1m_2$ and $W = (k\cdot p') -
m_1m_2$. From these we can deduce an important relation that becomes
vital during the calculations, namely $W-w=k\cdot(p'-p)=(k\cdot
q)=\frac{-q^{2}}{2}$. The $w$ and $W$ are displayed here only for the
$K $ and $K'$ integrals, for the rest of the integrals we have used
$W=(k\cdot p) -m_{1}m_{2}-\frac{q^{2}}{2}$, see~\cite{Thesis} for
derivations. The coefficients to these are long and tedious to write
down properly, however, if required, they can be obtained from us by
contacting us.

For the above integrals the following constraints for the
non-analytical terms can be verified directly on the mass-shell. We
will use ${}_{k}I\sim
\frac{1}{\ell^{2}(\ell+q)^{2}[(\ell+k)^{2}-m_{1}^{2}]}$ for $k$ and
${}_{p}I\sim \frac{1}{\ell^{2}(\ell+q)^{2}[(\ell-p)^{2}-m_{2}^{2}]}$
for $p$, ${}_{p'}I\sim
\frac{1}{\ell^{2}(\ell+q)^{2}[(\ell+p')^{2}-m_{2}^{2}]}$ for $p'$ and
no particular choice is needed for contractions with $q$
\begin{equation*}
K_{\mu\nu}\eta^{\mu\nu}=K'_{\mu\nu}\eta^{\mu\nu}=I_{\mu\nu\alpha}\eta^{\mu \nu } = I_{\mu\nu}\eta^{\mu\nu} = J_{\mu\nu}\eta^{\mu\nu} = 0
\end{equation*}
\begin{equation*} 
K_{\mu\nu}q^\mu = -\frac{q^2}{2}K_{\nu}, \ \ \ \ K_{\mu}q^\mu =
-\frac{q^2}{2}K, \ \ \ \ K'_{\mu\nu}q^\mu = -\frac{q^2}{2}K'_{\nu}, \ \
\ \ K'_{\nu}q^\nu = -\frac{q^2}{2}K'
\end{equation*}
\begin{equation*} 
K_{\mu\nu}p^\mu = -\frac{1}{2}{}_{k}I_{\nu}, \ \ \ \ K_{\mu}p^\mu =
-\frac{1}{2}{}_{k}I, \ \ \ \ K_{\mu\nu}k^\mu = \frac{1}{2}{}_{p}I_{\nu}, \ \
\ \ K_{\nu}k^\nu = \frac{1}{2}{}_{p}I
\end{equation*}
\begin{equation*} 
 K'_{\mu\nu}p'^\mu = \frac{1}{2}{}_{k}I_{\nu}, \ \ \ \ K'_{\nu}p'^\nu
= \frac{1}{2}{}_{k}I, \ \ \ \  K'_{\mu\nu}k^\mu = \frac{1}{2}{}_{p'}I_{\nu}, \ \ \ \ K'_{\nu}k^\nu
= \frac{1}{2}{}_{p'}I
\end{equation*}
\begin{equation*} 
I_{\mu\nu\alpha}q^\alpha = -\frac{q^2}{2}I_{\mu\nu}, \ \ \ \ I_{\mu\nu}q^\nu  = -\frac{q^2}{2}I_{\mu}, \ \ \ \   I_{\mu}q^\mu  = -\frac{q^2}{2}I \ \ \ \
J_{\mu\nu}q^\nu = -\frac{q^2}{2}J_{\mu}, \ \ \ \ J_{\mu}q^\mu   = -\frac{q^2}{2}J  
\end{equation*}
\begin{equation*}
{}_{k}I_{\mu\nu\alpha}k^\alpha = \frac{1}{2}J_{\mu\nu}, \ \ \ \ {}_{k}I_{\mu\nu}k^\nu  = \frac{1}{2}J_{\mu}, \ \ \ \   {}_{k}I_{\mu}k^\mu  = \frac{1}{2}J  
\end{equation*}
\begin{equation*}
{}_{p}I_{\mu\nu\alpha}p^\alpha = -\frac{1}{2}J_{\mu\nu}, \ \ \ \ {}_{p}I_{\mu\nu}p^\nu  = -\frac{1}{2}J_{\mu}, \ \ \ \   {}_{p}I_{\mu}p^\mu  = -\frac{1}{2}J  
\end{equation*}
There seems to be a typo in~\cite{Bohr1} the metric in
$I_{\mu\nu\alpha}\eta^{\mu\nu}$ is written as
$I_{\mu\nu\alpha}\eta^{\alpha\beta}$.
\twocolumngrid

\end{document}